\documentstyle[12pt]{article}
\newcommand{\nn}{\nonumber}
\newcommand{\bd}{\begin{document}}
\newcommand{\ed}{\end{document}}
\newcommand{\bc}{\begin{center}}
\newcommand{\ec}{\end{center}}
\newcommand{\be}{\begin{eqnarray}}
\newcommand{\ee}{\end{eqnarray}}
\newcommand{\ba}{\begin{array}}
\newcommand{\ea}{\end{array}}
\newcommand{\eqn}{\global\def\theequation}
\newcommand{\sw}{sin^2 \theta_W}
\newcommand{\fbd}{f_B}
\renewcommand{\thefootnote}{\alph{footnote}}
\newcommand{\se}{\section}
\newcommand{\sse}{\subsection}
\newcommand{\bi}{\bibitem}
\def\figcap{\section*{Figure Captions\markboth
     {FIGURECAPTIONS}{FIGURECAPTIONS}}\list
     {Figure \arabic{enumi}:\hfill}{\settowidth\labelwidth{Figure 999:}
     \leftmargin\labelwidth
     \advance\leftmargin\labelsep\usecounter{enumi}}}
\let\endfigcap\endlist \relax
\def\reflist{\section*{References\markboth
     {REFLIST}{REFLIST}}\list
     {[\arabic{enumi}]\hfill}{\settowidth\labelwidth{[999]}
     \leftmargin\labelwidth
     \advance\leftmargin\labelsep\usecounter{enumi}}}
\let\endreflist\endlist \relax

\begin{document}
\tolerance=10000
\baselineskip=7mm
\begin{titlepage}

 \vskip 0.5in
 \null
\begin{center}
 \vspace{.15in}
{\LARGE {\bf Study of $B_{s,d}\to l^+l^-\gamma$ Decays}
}\\
\vspace{1.0cm}
  \par
 \vskip 2.1em
 {\large
  \begin{tabular}[t]{c}
{\bf C.~Q.~Geng$^a$, C.~C.~Lih$^{a}$ and Wei-Min Zhang$^{b}$}
\\
\\
{\sl ${}^a$Department of Physics, National Tsing Hua University}
\\  {\sl  $\ $ Hsinchu, Taiwan, Republic of China }
\\
\\
and
\\
\\
       {\sl ${}^b$Department of Physics, National Cheng Kung University}
\\   {\sl  $\ $Tainan, Taiwan,  Republic of China }\\
   \end{tabular}}
 \par \vskip 5.3em
 {\Large\bf Abstract}
\end{center}

We study the decays of $B_{s,d} \to l^+ l^- \gamma\ (l=e, \mu,\tau )$
within
the light-front model.
We calculate the tensor type form factors and use these
form factors to evaluate the decay branching ratios.
We find that, in the standard model, the branching ratios of
$B_{s(d)} \to \l^+ \l^- \gamma$ ($\l=e, \mu,\tau$) are
$7.1\times 10^{-9}$ ($1.5\times 10^{-10}$),
$8.3\times 10^{-9}$ ($1.8\times 10^{-10}$),
$1.6\times 10^{-8}$ ($6.2\times 10^{-10}$), respectively.

\end{titlepage}

\se{Introduction}
$\ \ \ $

It is well known that $B$ physics is important to
determine the elements of the
Cabibbo-Kobayashi-Maskawa (CKM) matrix\cite{ckm}
and physics beyond the standard model.
Recently, the interest has been focused on the
rare $B$ meson decays induced by the flavor changing
neutral current (FCNC) due to the CLEO
measurement of the radiative $b \to s\gamma$
decay \cite{cleo}. In the standard model,
these rare decays occur at loop level and provide us
information on the parameters of the CKM matrix
elements as well as various hadronic form factors,
such as the $B$ meson decay constant $f_B$.

As in the decays of $B^+ \to l^{+}\nu_l$ ,
the helicity suppression effect is
also expected in the flavor
changing neutral current processes of $B_{s,d} \to l^{+}l^{-}$.
These decays are sensitive probes of top quark
couplings \cite{bsm} such as the CKM elements
$V_{ts(d)}$. The decay widths of
these leptonic decay modes are given by:
\be
\Gamma ( B_q \to l^+ l^-) =
\frac{\alpha^2 G_F^2 f_{B_q}^2 }{16 \pi^3}\,m_{B_q}^{3}
\left(\frac{m_l^2}{m_{B_q}^{2}}\right)
\vert V_{tb} V_{tq}^* \vert^2 C_{10}^2 \,,
\label{1}
\ee
where $G_{F}$ is the fermi constant, $M_{B_{q}}$ and $m_{l}$ are $B_{q}$
meson and lepton masses and $C_{10}$ is the Wilson coefficient.
Form Eq.(\ref{1}), one has that
$B(B_{s}\to e^{+}e^{-}, \mu^+\mu^-, \tau^+\tau^-)
\simeq( 6\times 10^{-8}, 2.6\times 10^{-1}, 1.0 )\times 10^{-6}$ and
$B(B_{d}\to e^{+}e^{-}, \mu^+\mu^-, \tau^+\tau^- )\simeq( 4.2\times 10^{-7},
1.8\times 10^{-2}, 4.5 )\times 10^{-8}$ by taking
$|V_{tb}V_{ts}^*|=0.04$, $|V_{tb}V_{td}^*|=0.01$,
$f_{B_{s,d}}\simeq 200\ MeV$, $\tau_{B_s}\simeq 1.61\ ps$ and
$\tau_{B_d}\simeq 1.5\ ps$ \cite{pdg}.
It is clear that the rates for the light lepton modes are too small to be
measured due to the helicity suppressions, while that for the $\tau$ channel,
although there is no suppression, it is hard to be
observed experimentally because of the low efficiency.

It has been pointed out that the radiative leptonic $B$ decays of
 $B^+\to \l^{+}\nu_l\gamma$ ($\l=e, \mu$) have larger decay rates than
that purely leptonic ones \cite{rad11,rad12,rad13,rad14,rad15,geng2,rad17,yan}.
Similar enhancements were also found for the
radiative decays
of $B_{s,d}\to \l^{+}\l^{-}\gamma$ in the constituent and light-cone sum
rule
quark models \cite{rad1,rad2}.
In fact, the decay amplitudes of $B_{s,d}\to \l^{+}\l^{-}\gamma$
can be divided into the ``internal-bremsstrahlung'' (IB) parts, where the
photon emits from the external charged leptons, which are still
helicity suppressed for the light charged
lepton modes, and the ``structure-dependent'' (SD) ones, in which
one of the photon comes from intermediate states of $B_q \to l^+l^-$,
which are free of the helicity suppression. Therefore, the decay rates of
$B_q\to l^{+}l^{-}\gamma$ ( $l=e, \mu$ ) might have an enhancement
with respect to the purely leptonic modes of $B_q\to l^{+}l^{-}$ if the SD
contributions to the decays are dominant.

In this paper, we will use the light front
quark model \cite{geng2,rad17,lf1,lf2,cheng,wmz} to evaluate
the hadronic matrix elements and study the decay rates for
$B_{s,d}\to l^{+}l^{-}\gamma$.
It is known that as the recoil
momentum increases, we have to start considering relativistic
effects seriously. The light front quark model \cite{wmz} is
the widely accepted relativistic quark model in which a
consistent and relativistic treatment of quark spins and the
center-of-mass motion can be carried out. In this work, we
calculate for the first time the tensor form factors in
$P\to\gamma$ ($P$ is a pseudoscalar meson) directly at time-like
momentum transfers by using the relativistic light-front
hadronic wave function. Within the framework of light-front
quark model, one can calculate in the frame where the
momentum transfer is purely longitudinal, $i.e$, $p_{\bot}=0$ and
$p^2=p^+p^-$, which covers the entire range. Thus, we
give their dependence on the momentum transfer $p^{2}$ in
whole kinematic region of $0\leq p^{2}\leq
p_{\max }^{2}$.

The paper is organized as follows. In Sec.~2,
we present the decay amplitudes of $B_{s,d} \to
l^{+}l^{-}\gamma$. We study
the form factors in the $B\to\gamma$ transition
within the light front framework in Sec.~3.
In Sec.~4, we calculate the decay branching ratios.
We also compare our results with
those in literature \cite{rad1,rad2,radtau}.
We give our conclusions in Sec.~5.

\se{Decay Amplitudes}

\ \ \

 The main contributions for the processes of
 $B_q \to l^{+}l^{-}\gamma$ ($l=e, \mu$, $\tau$)
arise from the effective Hamiltonian that induces
the purely leptonic modes of $B_q \to l^{+}l^{-}$.
The QCD-corrected amplitude for $b \to l^{+}l^{-}q$
in the SM is given by \cite{c1,c2}:
\be
{\cal M}&=&-\frac{G_{F}\alpha}{\sqrt{2}\pi}V_{tb}V_{tq}^{*}
\Bigg\{C_{9}^{eff}(\bar{q}\gamma_{\mu}P_{L}b)\bar{l}\gamma^{\mu}l+C_{10}
\bar{q}\gamma_{\mu}P_{L}b\bar{l}\gamma^{\mu}\gamma_{5}l
 \nn\\
 && ~~~~~~~~
-2\frac{C_7}{p^2}\bar{q}i\sigma_{\mu\nu}p^{\nu}(m_{b}P_{R}+m_{q}P_{L})b
\bar{l}\gamma^{\mu}l\Bigg\} ,
\label{2}
\ee
where $q=d$ or $s$, $P_{L,R}$ = $(1\mp \gamma_5)/2$,
and $p$ is the momentum transfer, which
is equal to the momentum of the lepton pair.
The Wilson coefficients of $C_7, C_{9}^{eff}$ and $C_{10}$
can be found, for example, in \cite{c1,c2,c4,c3}, respectively.

The amplitude for $B_{q}\to l^{+}l^{-}\gamma $ can be written as:
\be
{\cal M}(B_{q}\to l^{+}l^{-}\gamma)={\cal M}_{IB}+{\cal M}_{SD}
\label{2.1}
\ee
where ${\cal M}_{IB}$ and ${\cal M}_{SD}$ represent the IB
and SD contributions, respectively.
For the IB part, the amplitude is clearly proportional to the
lepton mass $m_l$ and it is found to be
\be
{\cal M}_{IB}&=&-ie\frac{G_{F}\alpha}{\sqrt{2}\pi}V_{tb}V_{tq}^{*}f_{B_q}
C_{10}m_{\l}
\left[\bar{\l}(\frac{\not{\! }\epsilon\not{\! }P_{B}}{2p_{1}\cdot
q_{\gamma}
}
-\frac{\not{\! }P_{B}\not{\! }\epsilon}{2p_{2}\cdot q_{\gamma}
})\gamma_{5}\l\right] ,
\label{3.1}
\ee
where $P_B$, $p_1$, $p_2$ and $q_{\gamma}$, are the momentum of $B_q$,
$\l^+$, $\l^-$ and $\gamma$,
respectively. In Eq.(4),
the $f_{B_q}$ is the $B_q$ meson decay constant, defined by:
\be
<0|\bar{s}\gamma^{\mu}\gamma_5 b|B_q(p)> &=&-if_{B_q}p^{\mu}.
\ee
When a photon emitted from the charged internal line,
the amplitude is suppressed by a factor $m_{b}^{2}/m_{W}^{2}$
in the Wilson coefficients and the main contribution comes
from the photon radiating from the
initial quark line. Therefore ${\cal M}_{SD}$ can be written as
\be
{\cal M}_{SD}&=&-\frac{G_{F}\alpha}{\sqrt{2}\pi}V_{tb}V_{tq}^{*}
\Bigg\{C_{9}^{eff}\langle\gamma(q_{\gamma}
)|\bar{q}\gamma_{\mu}P_{L}b|B(p+q_{\gamma}
)\rangle
\bar{l}\gamma^{\mu}l
 \nn\\
 && ~~~~~~~~
+C_{10}
\langle\gamma(q_{\gamma}
)|\bar{q}\gamma_{\mu}P_{L}b|B(p+q_{\gamma}
)\rangle\bar{l}\gamma^{\mu}\gamma_{5}l
 \nn\\
 && ~~~~~~~~
-2\frac{C_7m_b}{p^2}
\langle\gamma(q_{\gamma}
)|\bar{q}i\sigma_{\mu\nu}p^{\nu}P_R b|B(p+q_{\gamma}
)\rangle
\bar{l}\gamma^{\mu}l\Bigg\} .
\label{3.2}
\ee
From the amplitude in Eq.(6),
we see that to find the decay rates,
one has to evaluate the hadronic matrix elements.
These elements can be parameterized as follows:
\be
\langle\gamma (q_{\gamma}
)|\bar{q}\gamma_{\mu }\gamma _{5}b|B_{q}(p+q_{\gamma}
)\rangle &=&
-e{\frac{F_{VA}}{%
M_{B_{q}}}}\left[ (p\cdot q_{\gamma}) \epsilon ^*_\mu
-(\epsilon ^{*}\cdot p)q_{\gamma \mu }\right] ,
\nonumber\\
\langle\gamma (q_{\gamma}
)|\bar{q}\gamma_{\mu }b|B_{q}(p+q_{\gamma}
)\rangle &=&
ie{\frac{F_{VV}}{M_{B_{q}}}}\varepsilon_{\mu \alpha \beta \nu }
\epsilon^{*\alpha }p^{\beta }q_\gamma^\nu \, ,
\label{4}
\ee
\be
\langle\gamma(q_{\gamma}
)|\bar{q}i\sigma_{\mu\nu}p^{\nu}\gamma_5b|B_{q}(p+q_{\gamma}
)\rangle &=&
ieF_{TA}\left[(p \cdot q_{\gamma})\epsilon^*_{\mu}
-(\epsilon^* \cdot p) {q_{\gamma}}_\mu)\right] , \nonumber\\
\langle\gamma(q_{\gamma}
)|\bar{q}i\sigma_{\mu\nu}p^{\nu}b|B_{q}(p+q_{\gamma}
)\rangle &=&
eF_{TV}\varepsilon_{
\mu\nu\alpha\beta}\epsilon^{*\nu} q_{\gamma}^\alpha p^\beta ,
\label{5}
\ee
where the $\epsilon _{\mu }$ is the photon polarization
vector, $q_{\gamma}$ and $p+q_{\gamma}$  are the four momenta
of the photon and the $B_{q}$ meson, and
$F_{VA}$, $F_{VV}$, $F_{TA}$ and $F_{TV}$ are the form factors of
axial-vector, vector, axial-tensor and tensor, respectively.
Form Eqs.(7) and (8), we rewrite the amplitude of Eq.(6) as,
\be
{\cal M}_{SD} &=& \frac{\alpha G_F}{2\sqrt{2}\pi}V_{tb}V_{ts}^*
\Bigg{\{} \epsilon_{\mu \nu\alpha \beta} {\epsilon^*}^\nu p^\alpha
{q_{\gamma}}^\beta \left[ A \, \bar \l \gamma^\mu \l + C \, \bar \l
\gamma^\mu
\gamma_5 \l \right] ~ \nn \\
&& +~  i \left[ (p\cdot q_{\gamma}) \epsilon_\mu^*
- (\epsilon^* \cdot p ) {q_{\gamma}}_\mu \right]
\left[ B \, \bar \l \gamma^\mu \l + D  \, \bar \l \gamma^\mu \gamma_5
\l \right] \Bigg{\}}~,
\label{5.1}
\ee
where the factors of $A$-$D$ are defined by
\be
A &=& \frac{C_9^{eff}}{M_{B}} F_{VA}(p^2)-2\,C_7\frac{m_b}{p^{2}}F_{TA}(p^2)~, \nn \\
B &=& \frac{C_9^{eff}}{M_{B}} F_{VV}(p^2)-2\,C_7\frac{m_b}{p^{2}}F_{TV}(p^2)~, \nn \\
C &=& \frac{C_{10}}{M_{B}} F_{VA}(p^2), \nn \\
D &=& \frac{C_{10}}{M_{B}} F_{VV}(p^2),
\ee
respectively. The form factors of $F_{VA}$ and $F_{VV}$
have been evaluated previously in the
light-front model \cite{geng2}, while that of
$F_{TA}$ and $F_{TV}$ shall be studied in the next section.

\section{Form Factors in the Light Front Model}

\ \ \

In this section, we will use the light-front approach to calculate
 the tensor type form factors in Eq.~(\ref{5}) for $B_{q} \to \gamma$
($q=s$ or $d$)
transition. In this approach, the $B$ meson bound state consists
of an anti-quark $\bar{b}$ and a
quark $q$ with the total momentum of $(p+q_{\gamma})$ and it is given by
\begin{eqnarray}
|B(p+q_{\gamma}
)>&=& \sum_{\lambda _{1}\lambda_{2}}\int [dk_{1}][dk_{2}]
2(2\pi)^{3}\delta ^{3}(p+q_{\gamma}-k_{1}-k_{2})  \nonumber \\
&& ~~~~~~~~ \times \Phi _{B}^{\lambda _{1}\lambda _{2}}(x,k_{\bot})
b_{\bar{b}}^{+}(k_{1},\lambda _{1}) d_{q}^{+}( k_{2},\lambda _{2})
|0>\,,
\end{eqnarray}
where $k_{1(2)}$ is the on-mass shell light front momentum of the internal
quark $\bar{b}(q)$, the light front relative momentum variables $(x,k_{\bot
})$ are defined by
\begin{eqnarray}
k_1^+= x(p+q_{\gamma}
)^{+}\,,\ k_{1\bot} = x(p+q_{\gamma}
)_{\bot}+k_{\bot}\,,
\end{eqnarray}
and
\be
\Phi _{B}^{\lambda _{1}\lambda _{2}}(x,k_{\bot })=\left( \frac{%
2k_{1}^{+}k_{2}^{+}}{M_{0}^{2}-\left( m_{q}-m_{b} \right) ^{2}}\right)^{%
\frac{1}{2}}\overline{u}\left( k_{1}, \lambda _{1}\right) \gamma^{5}v\left(
k_{2},\lambda _{2}\right) \phi(x,k_{\bot}) \,,  \label{n6}
\ee
with $\phi(x,k_{\bot})$ being the momentum distribution amplitude.
The amplitude of $\phi(x,k_{\bot})$ can be solved in principle by
the light-front QCD bound state equation\cite{wmz2,wmz3}.
Here, we use the Gaussian type wave function:
\be
\phi(x,k_{\bot})=N\sqrt{\frac{dk_{z}}{dx}}
\exp \left( -\frac{\vec{k}^{2}} {2\omega_{B}^{2}}\right) \,,
\label{7}
\ee
where
\begin{eqnarray}
& & [dk_1]= {\frac{dk^+dk_{\bot}}{2(2\pi)^3}}\, , \ \ N = 4 \left({\frac{\pi%
}{\omega_{B}^{2}}}\right)^{\frac{3}{4}}\, ,  \nonumber \\
& & k_{z} =\left( x-\frac{1}{2}\right) M_{0}+\frac{m_{b}^{2}-m_{q}^{2}}{%
2M_{0}} \, , \ \ M_0^2={\frac{k^2_{\bot}+m_q^2}{x}}+{\frac{k^2_{\bot}+m_b^2}{%
1-x}} \, ,  \nonumber \\
& & \sum_\lambda u(k,\lambda) \overline{u}(k,\lambda) = {\frac{m + \not{\!
}k }{k^+}} \, , \ \ \sum_\lambda v(k,\lambda) \overline{v}(k,\lambda) = - {%
\frac{m - \not{\! }k }{k^+}} \, .
\end{eqnarray}
We note that the wave function in Eq.(14) could be also applied to
many other of hadronic transitions. For the
gauged photon state, one has \cite{geng2}
\begin{eqnarray}
|\gamma (q_{\gamma}
)> &=& N^{\prime}\Bigg\{a^{+}(q_{\gamma},\lambda) + \sum_{\lambda_{1}
\lambda_{2}}\int [dk_{1}][dk_{2}]2(2\pi)^{3}\delta^{3} (q_{\gamma}
-k_{1}-k_{2})
\nonumber \\
&& ~~~~~~~~~ \times \Phi _{q\bar{q}}^{\lambda_{1}\lambda_{2}\lambda}
(q_{\gamma}, k_{1},k_{2}) b_{q}^{+}(k_{1},\lambda _{1}) d_{\bar{q}}^{+}
(k_{1},\lambda
_{2}) \Bigg\} | 0 > \, ,
\end{eqnarray}
where
\begin{eqnarray}
\Phi_{q\bar{q}}^{\lambda _{3}\lambda _{4}\lambda }
(q_{\gamma},k_{1},k_{2}) &=&\frac{%
e_q}{ED}\chi _{-\lambda _{2}}^{+} \left\{-2\frac{
{q_{\gamma}}_{\bot }\cdot \epsilon
_{\bot }} {q_{\gamma}^{+}}-\tilde{\sigma}_{\bot}\cdot
\epsilon_{\bot } \frac{\tilde{\sigma}_{\bot
}\cdot k_{2_{\bot }}+im_{2}}{k_{2}^{+}} \right.  \nonumber \\
&& ~~~~~~~ \left.-\frac{\tilde{\sigma}_{\bot }\cdot
k_{1_{\bot }}+im_{1}} {k_{1}^{+}} \tilde{\sigma}_{\bot }
\cdot \epsilon _{\bot }\right\} \chi _{\lambda_{1}}\,,
\label{pqq}
\end{eqnarray}
with
\begin{eqnarray}
ED&=&\frac{{q_{\gamma}}_{\bot
}^{2}}{q_{\gamma}^{+}}-\frac{k_{1_{\bot}}^{2}+m_{1}^{2}}
{k_{1}^{+}%
}-\frac{k_{2_{\bot }}^{2}+m_{2}^{2}}{k_{2}^{+}} \,.
\end{eqnarray}
In Eq.~({\ref{pqq}), we have used the two-component form of
the light-front quark fields \cite{wmz4,wmz}.
Since the physically accessible kinematic region is
$0\leq p^{2}\leq p_{\max }^{2}$ where $p_{max}^2 = M_B^2$,
to calculate the matrix elements in Eq.~(\ref{5}), we choose a frame
where the transverse momentum $p_{\bot}$ = $0$. Then $p^{2}=p^{+}p^{-}
\geq 0$ which can cover the entire range of the momentum transfers.
By considering the ``good" component of $\mu=+$,
the tensor current in Eq.~(\ref{5}) can be rewritten as:
\begin{eqnarray}
<\gamma (q_{\gamma}
)|(q_{+}^{+}\gamma^{0}\gamma_5b_{-}-
q_{-}^{+}\gamma^{0}\gamma_5b_{+})|B(p+q_{\gamma})> &=&eF_{TA}%
\left( \epsilon _{\bot }^{*}\cdot {q_{\gamma}}_{\bot }\right) \,,
\nonumber \\
<\gamma (q_{\gamma})|(q_{+}^{+}\gamma^{0}b_{-}-
q_{-}^{+}\gamma^{0}b_{+})|B(p+q_{\gamma}
)>&=&-ieF_{TV} \epsilon ^{ij}\epsilon
_{i}^{*}{q_{\gamma}}_{j}\,,  \label{ff}
\end{eqnarray}
where $q_{+}(b_{+})$ and $q_{-}(b_{-})$ are the light-front up
and down component of the quark fields. In the two-component
form \cite{wmz4,wmz}, they are expressed by
\be
q_{+}&=&\left(\ba{c}\chi \\ 0\ea\right) \,
\ee
and
\be
q_{-}&=&\frac{1}{i\partial^{+}}(i\alpha_{\bot}\cdot
\partial_{\bot}+\beta m_{q})q_{+}
  = \left(\ba{c}0 \\ \frac{1}{\partial^{+}}
(\tilde{\sigma}_{\bot} \cdot\partial_{\bot}+m_{q})
\chi_{q}\ea\right) \, .
\ee
In Eq.~(21), $\chi_{q}$ is a two-component spinor field
and $\sigma$ is the Pauli matrix. The form factors of
$F_{TA}$ and $F_{TV}$ in Eqs. (\ref{ff}) are then found to be:
\begin{eqnarray}
F_{TA}(p^{2}) &=&\int \frac{dx\,d^{2}k_{\bot }}{2(2\pi)^{3}}%
\Phi \left( x',k_{\bot }^{2}\right) \nonumber
\\
&&~~~~~~~~ \times \left\{ \frac{1}{3}\frac{A_1+A_2\, k_{\bot }^{2} \Theta}{%
m_{b}^{2}+ k_{\bot}^{2}}+\frac{1}{3}\frac{B_1+B_2\, k_{\bot }^{2}\Theta } {%
m_{q}^{2}+k_{\bot }^{2}} \right\}\,  \label{fffa}
\end{eqnarray}
and
\begin{eqnarray}
F_{TV}(p^{2}) &=&-\int \frac{dx\,d^{2}k_{\bot }}{2\left( 2\pi
\right) ^{3}}\Phi \left( x',k_{\bot }^{2}\right)
\nonumber \\
&&~~~~~~~~ \times \left\{ \frac{1}{3}\frac{C_1+C_2\, k_{\bot }^{2} \Theta }{
m_{b}^{2}+k_{\bot }^{2}}\ +\frac{1}{3}\frac{D_1+ D_2\, k_{\bot }^{2}\Theta }
{m_{q}^{2} +k_{\bot }^{2}}\right\}\,
\label{fffv}
\end{eqnarray}
where
\be
A_1 &=& \frac{2}{xx'^{2}(1-x')(1-x)}\Bigg\{
(x'+x-2x'x)\left[x'(x-1)-x(2x-1)\right]k_{\bot}^{2}  \nonumber \\
&&+x\left[(x-x')+2x'x(1-x)\right]m_{b}^{2}
+2x^{2}(1-x')^{2}m_{b}m_{q}\Bigg\}\,,  \nonumber \\
A_2 &=& \frac{2(x-x')}{xx'^{2}(1-x')(1-x)}\Bigg\{
(x'+x-2x'x)(1-2x)k^{2}_\bot  \nonumber \\
&&+2x(1-x')m_{q}m_{b}-x(1-2x)(1-x')^{2}m_{q}^{2}
+x'(1+x'x-2xx'^{2})m_{b}^{2}\Bigg\}\,,  \nonumber \\
B_1 &=& \frac{2}{x'x(1-x)(1-x')^{2}}\Bigg\{
(x'+x-2x'x)(1-2x+2x^{2}-x'x)k_{\bot}^{2}  \nn \\
&&+2xx'(1-x')(1-x)m_q\,m_b+(1-x')
(x'+x-4x'x+2x'x^{2})m_{q}^{2} \Bigg\}\,,  \nn \\
B_2 &=& \frac{2(x-x')}{x'x(1-x)(1-x')^{2}}\Bigg\{
-(x'+x-2x'x)(1-2x)k_{\bot}^{2}
\nn \\
&&-\left[(1-2x)x'^{2}(1-x)m_{b}^{2}
+(1-x')(x'+x-3x'x-2x^{2}(1-x'))m_{q}^{2}\right]\Bigg\}\,,  \nn \\
C_1 &=& \frac{2}{x{x'}^2(1-x')(1-x)}\Bigg\{
(x-x'+xx')(x'+x-2x'x)k_{\bot}^{2}  \nn \\
&&-2x^{2}(1-x')^{2}m_bm_q
+x'\left[(x-x')-2(1-x')x^{2}\right]m_{b}^{2}\Bigg\}\,, \nonumber \\
C_2 &=& \frac{2(x-x')}{x{x'}^2(1-x')(1-x)}\Bigg\{
(x'+x-2x'x) k_{\bot}^{2}+x'(1-2x+xx')m_{b}^{2}\nn \\
&& -m_{q}^{2}x(1-x')^{2}-2x(1-x')m_q\,m_b\Bigg\}\,, \nonumber \\
D_1 &=& \frac{2}{x(1-x)x'(1-x')^{2}}\Bigg\{
-(1-x)(1-2x+x')(x'+x-2x'x)k_{\bot}^{2}  \nn \\
&&-\left[(1-x')(x'+x-2x^{2}-2x'x+2x^{2}x')m_{q}^{2}
+ 2{x'}^2(1-x)^2 m_{q}m_{b}\right]\Bigg\}\,,  \nonumber \\
D_2 &=& \frac{2(x-x')}{x'(1-x)x(1-x')^{2}}\Bigg\{
(x'+x-2x'x) k_{\bot}^{2} +x'^{2}(1-x)m_{b}^{2} \nn \\
&&+(1-x')(x'-x-x'x)m_{q}^{2}\Bigg\}\,, \nonumber \\
\Phi (x,k_{\bot}^2) &=& N\left( {\frac{2x(1-x) }{M_0^2-(m_q-m_b)^2}}%
\right)^{1/2} \sqrt{{\frac{dk_{z}}{dx}}}\exp \left( -{\frac{\vec{k}^{2}}{%
2\omega_B^2}}\right)\,,  \nonumber \\
\Theta &=& {\frac{1}{\Phi(x,k_{\bot}^2) }} {\frac{d\Phi(x,k_{\bot}^{2})}{%
dk_{\bot}^2}} \, ,  \nonumber \\
x^{\prime}&=&x\left(1-{\frac{p^2}{M_B^2}}\right),\
\vec{k}=(\vec{k}_{\bot}, \vec{k}_{z}) \,.
\label{ffff}
\ee

To illustrate numerical results, we input $m_{q}=m_{s}=0.4,$
$m_{b}=4.5$ in $GeV$, and $\omega =0.55$ $GeV$ which is
determined by fitting $f_{B_s}=200~GeV$ via Eq.~(5).
The values of $F_{TA}$ and $F_{TV}$ in the entire
range of $p^2$ are shown in Fig. 2.
We note that the tails of $F_{TV,TA}$ at the large
momentum transfers go down may be the
long distance contribution associated with $B-B^*-\gamma$
vertex, which is not included in the present work.
It is interesting to note that the formulas in
Eqs.~(\ref{fffa}) and (\ref{fffv}) can be used for
other pseudoscalars, such as pions and kaons,
to the photon transitions as
well once we put in the corresponding masses.

\section{Decay Branching Ratios}

\ \ \

The partial decay width for $B_q \to l^+ l^- \gamma$
in the $B$ rest frame is given by
\be
d\Gamma={1\over 2M_B}|{\cal M}|^2
(2\pi)^4\delta^4(P_B-p_{1}-p_{2}-q_{\gamma})
{d\vec{q_{\gamma}}\over (2\pi)^3 2E_{\gamma
}} {d\vec{p}_{1}\over (2\pi)^3
2E_{1}} {d\vec{p}_{2}\over (2\pi)^3 2E_{2}},
\ee
where the square of the matrix element can be written as
\be
|{\cal M}|^2=|{\cal M}_{IB}|^2+|{\cal M}_{SD}|^2+2Re({\cal M}_{IB}{\cal
M}_{SD}^{*})\,.
\ee
To describe the kinematic of $B_q \to l^+ l^- \gamma$, we defined
two variables of
$x_{\gamma}=2P_{B}\cdot q_{\gamma}
/M_B$ and $y=2P_{B}\cdot p_{1}/M_B$.
 One can easily write the transfer momentum $p^2$ in term of
$x_{\gamma}$ as
\be
p^2=M_{B}^{2}(1-x_{\gamma}).
\ee
The double differential decay rate is found to be
\be
\frac{d^{2}\Gamma ^{l}}{dx_{\gamma}d\lambda } &=&\frac{M_{B}}{256\pi
^{3}}\left| M\right| ^{2}=C\rho (x_{\gamma},\lambda) ,
\ee
where
\be
C=\alpha|\frac{\alpha V_{tb}V_{ts}^{*}}{8\pi^{2}}|^{2}G_{F}^{2}M_{B}^{5}\,,
\ee
and
\be
\rho (x_{\gamma},\lambda) &=&\rho_{IB}(x_{\gamma},\lambda)
+\rho_{SD}(x_{\gamma},\lambda)+\rho_{IN}(x_{\gamma},\lambda),
\ee
with
\be
\rho_{IB} &=&4|f_{B}C_{10}|^{2}\frac{r_l}{M_{B}^{2}
x_{\gamma}^{2}}\Bigg\{
\frac{x_{\gamma}^{2}-2x_{\gamma}+2-4r_{l}}{\lambda(1-\lambda)}-2r_{l}(
\frac{1}{\lambda^{2}}+\frac{1}{(1-\lambda)^{2}})\Bigg\},  \nonumber \\
\rho_{SD} &=&\frac{M_{B}^{2}}{8}x_{\gamma}^{2}\Bigg\{(|A|^{2}+|B|^{2})
\left[(1-x_{\gamma}+2r_{l})
-2(1-x_{\gamma})(\lambda-\lambda ^{2})\right]  \nn \\
&&+(|C|^{2}+|D|^{2})\left[(1-x_{\gamma}-2r_{l})
-2(1-x_{\gamma})(\lambda-\lambda ^{2})\right]  \nn \\
&&+2Re(B^*C+A^*D)
(1-x_{\gamma})(2\lambda-1)\Bigg\} ,  \nonumber \\
\rho_{IN} &=&f_{B}C_{10}r_{l}\Bigg\{Re(A)
\frac{x_{\gamma}}{\lambda(1-\lambda)}+Re(D)
\frac{x_{\gamma}(1-2\lambda)}{\lambda(1-\lambda)}\Bigg\}\, .
\ee
Here $\lambda =(x_{\gamma}+y-1)/x_{\gamma}$ and $r_l=m_{l}^2/M_{B}^2$ and
the physical regions for $x_{\gamma}$ and $\lambda$ are given by:
\be
0 &\leq &x_{\gamma}\leq 1-4r_{l} \,,  \nonumber \\
\nonumber \\
\frac{1}{2}-\frac{1}{2}\sqrt{1-\frac{4r_{l}}{1-x_{\gamma}}} &\leq &\lambda
\leq \frac{1}{2}+\frac{1}{2}\sqrt{1-\frac{4r_{l}}{1-x_{\gamma}}}\,.
\ee

For the Wilson coefficients $C_7$ and $C_{10}$, we use the results given by
Refs.\cite{c1,c2} and they are
\be
C_7 = -0.315~~,~~C_{10} = - 4.642~. \nn
\ee
The analytic expressions for $C_9^{eff}$ in
the next-to-leading order approximation is given by\cite{c3}:
\be
C_9^{eff} &=& C_9 + 0.124 w(s) + g(\hat{m_c},s)
( 3 C_1 + C_2 + 3 C_3 + C_4 + 3 C_5 + C_6) \nn \\
&&- \frac{1}{2}g(\hat{m_q},s)(C_3 + 3 C_4)
- \frac{1}{2} g(\hat{m_b},s)(4 C_3 + 4 C_4 + 3 C_5 + C_6)  \nn \\
&& + \frac{2}{9} (3 C_3 + C_4 + 3 C_5 + C_6) ,
\label{18}
\ee
with
\be
C_9&=&4.227~,  \nn \\
( 3 C_1 + C_2 + 3 C_3 + C_4 + 3 C_5 + C_6)&=&0.359~,  \nn \\
(C_3 + 3 C_4)&=&-0.0659~,  \nn \\
(4 C_3 + 4 C_4 + 3 C_5 + C_6)&=&-0.0675~,  \nn \\
(3 C_3 + C_4 + 3 C_5 + C_6)&=&-0.00157~,
\label{19}
\ee
and $s=p^2/m_b^2$.
In the Eq.(\ref{18}), the function of $w(\hat{s})$ comes from the single-gluon
correction to the matrix element of $O_9$ and its form can be found in Refs.\cite{c2,c4},
while that of $g(\hat{m_i},s)$ from the one loop
contributions of the four quark operators $O_1$ - $O_6$,
given by \cite{c2,c3}:
\be
g(\hat{m_i},s)&=& - \frac{8}{9} \ln \hat{m_i} + \frac{8}{27} +
\frac{4}{9} y_i - \frac{2}{9} (2 + y_i) \sqrt{|1-y_i|}  \nn \\
&& \times \left\{\ba{cr}
(\ln \frac{1+\sqrt{1-y_i}}{1-\sqrt{1-y_i}} - i\pi )
&\quad\mbox{for \,\, $y_i<1$}\\
2\arctan \frac{1}{\sqrt{y_i-1}}
&\quad\mbox{for \,\, $y_i>1$} \ea \right.
\label{21}
\ee
with $y_i = \hat{m_i}^2 / s$ and $\hat{m_i} = m_i/m_b$.
By taking into account the
long distance effects mainly due to the $J/\psi$ family resonances, one may use
the replacement\cite{ld}
\be
g(\hat{m_c},s) \to g(\hat{m_c},s) - \frac{3 \pi}{\alpha^2}
\sum_{V=J/\psi , \psi'} \frac{\hat{m_V} Br( V \to l^+ l^-)
\hat{\Gamma}_{tot}^V}{s - \hat{m_V}^2 + i \hat{m_V}
\hat{\Gamma}_{tot}^V}
\ee
where $\hat{m_V} = m_V/m_b$ and $\hat{\Gamma}_{tot} = \Gamma /m_b$ .
The masses and decay widths of the corresponding mesons in Eq.(35)
are listed\cite{pdg} in table 1.

\begin{table}[h]\caption{Charmonium mass and width}
\begin{center}
\begin{tabular}{|c|c|c|c|}
\hline
Meson & Mass ($GeV$) & $Br(V \to l^+l^-)$ & $\Gamma_{total} (MeV)$  \\
\hline
$J/\psi$ & 3.097 & $6.0 \times 10^{-2}$ & 0.088 \\ \hline
$\psi$ & 3.686 & $8.3 \times 10^{-3}$ & 0.277 \\ \hline
\end{tabular}\end{center}\end{table}

In Figs.2 and 3 we present the differential decay rates of $B_s \to \mu^+
\mu^- \gamma$ and $B_s \to \tau^+\tau^- \gamma$ as functions of $x_{\gamma}$,
with and without resonance ($J/\psi~ and~ \psi'$) contributions. 
From these figures we see that the contributions from the IB parts,
corresponding to small $x_{\gamma}$ region, are infrared divergence.
To obtain the decay width of $B_s \to \tau^+\tau^- \gamma$,
a cut on the photon energy is needed.
Our results for the integrated
branching ratios without and with long-distance effects
as well as those given by the
constituent quark model \cite{rad2} and the light-cone QCD sum rule
model \cite{rad1,radtau}
are summarized
in tables 2 and 3, respectively.
Here, we have used the cut value of $\delta =0.01$ and
$m_c=1.5\ GeV$, $m_s=0.4\ GeV$, $\vert V_{tb}V_{ts}^*\vert = 0.045$,
$\vert V_{tb} V_{td}^*\vert = 0.01$, $\tau(B_s) = 1.61 \times 10^{-12}~s$ and
$\tau(B_d) = 1.5 \times 10^{-12}~s$\cite{pdg}.

\begin{table}[h]\caption{Integrated branching ratios for the radiative
leptonic $B_{s,d}$ decays without long-distance}
\begin{center}
\begin{tabular}{|c|c|c|c|c|c|c|}
\hline
Integrated Decay  &  &&&&Ref.&Ref.\\
Branching Ratios & IB & SD & IN & Sum  & 
\cite{rad2} &
 \cite{rad1,radtau}  \\ \hline
\hline
$10^{9}B(B_s\to e^{+}e^{-}\gamma)$ & $3.1\times 10^{-5}$ & $3.2$ &
$5.2\times 10^{-6}$ & $3.2$ &$6.2$ &$2.35$ \\ \hline
$10^{9}B(B_s\to \mu^+\mu^-\gamma)$ & $5.5\times 10^{-1}$ & $3.2$ &
$8.2\times 10^{-2}$ & $3.8$ &$4.6$ &$1.9$ \\ \hline
$10^{9}B(B_s\to \tau^+\tau^-\gamma)$ & & &  &  && \\
(Cut $\delta=0.01$) & $13.0$ &
$0.6$ & $6.4 \times 10^{-1}$ & $14.2$ &$-$& $9.54$ \\ \hline
$10^{10}B(B_d\to e^{+}e^{-}\gamma)$ & $1.4\times 10^{-5}$ & $1.5$ &
$2.4\times 10^{-6}$ & $1.5$ &$8.2$ &$1.5$ \\ \hline
$10^{10}B(B_d\to \mu^+\mu^-\gamma)$ & $2.6\times 10^{-1}$ & $1.5$ &
$3.8\times 10^{-2}$ & $1.8$ &$6.2$ &$1.2$ \\ \hline
$10^{10}B(B_d\to \tau^+\tau^-\gamma)$ & & &  &  && \\
(Cut $\delta=0.01$) & $5.7$ &
$2.4\times 10^{-1}$ & $2.7\times 10^{-1}$ & $6.1$ &$-$& $-$ \\ \hline
\end{tabular}\end{center}\end{table} 

\begin{table}[h]\caption{Integrated branching ratios for the radiative
leptonic $B_{s,d}$ decays with long-distance}
\begin{center}
\begin{tabular}{|c|c|c|c|c|c|}
\hline Integrated Decay  &  &&&&Ref.\\ Branching Ratios & IB
& SD & IN & Sum  & \cite{radtau} \\ \hline
\hline $10^{9}B(B_s\to e^{+}e^{-}\gamma)$ & $3.1\times 10^{-5}$ &
$7.1$ & $5.8\times 10^{-6}$ & $7.1$ &$-$ \\ \hline
$10^{9}B(B_s\to \mu^+\mu^-\gamma)$ & $5.5\times 10^{-1}$ & $7.1$ &
$9.4\times 10^{-2}$ & $8.3$ &$-$ \\ \hline
$10^{9}B(B_s\to \tau^+\tau^-\gamma)$ & & &  &&  \\ (Cut
$\delta=0.01$) & $13.0$ & $1.9$ & $7.9\times 10^{-1}$ & $15.7$ &$15.2$ \\
\hline $10^{10}B(B_d\to e^{+}e^{-}\gamma)$ & $1.4\times 10^{-5}$
& $1.5$ & $2.4\times 10^{-6}$ & $1.5$ &$-$ \\ \hline
$10^{10}B(B_d\to \mu^+\mu^-\gamma)$ & $2.6\times 10^{-1}$ & $1.5$
& $3.8\times 10^{-2}$ & $1.8$ &$-$ \\ \hline
$10^{10}B(B_d\to \tau^+\tau^-\gamma)$ & & &  &&  \\ (Cut
$\delta=0.01$) & $5.7$ & $2.3\times 10^{-1}$ & $2.5\times 10^{-1}$ & $6.2$ &$-$ \\
\hline
\end{tabular}\end{center}\end{table} 
We now compare our results with those in the literature
\cite{rad1,rad2,radtau}.
As shown in Table 2,
the decay branching ratios
for $B_{s} \to l^+ l^- \gamma$ ($\l=e, \mu$)
without long distance contributions
found in our calculations
are smaller than those in the constituent quark model \cite{rad2}, whereas
for $B_{d} \to l^+ l^- \gamma$ the statement are much smaller than ones.
It is mainly due to that in the constituent quark model
\cite{rad2} the decay rate
of $B_{q} \to l^+ l^- \gamma$ is proportional to the inverse of
the quark mass $m_q$.
It is interesting to see that our results are close to those in 
the light-cone QCD sum rule model \cite{rad1,radtau}. 
We note that the SD contributions for the decays in both constituent quark
and light-cone QCD sum rule models are sensitive
to the values of the decay constants $f_{B_{s,d}}$.

\section{Conclusions}

\ \ \

We have studied the decays of $B_{s,d} \to l^+ l^- \gamma$ within the
light-front model.
We have calculated the tensor type form factors and used these
form factors to evaluate the decay branching ratios.
We have found that, in the standard model, the branching ratios of 
$B_{s(d)} \to e^+ e^- \gamma$, $B_{s(d)} \to \mu^+ \mu^- \gamma$ and 
$B_{s(d)} \to \tau^+\tau^- \gamma$ are 
$7.1\times 10^{-9}$ ($1.5\times 10^{-10}$), 
$8.3\times 10^{-9}$ ($1.8\times 10^{-10}$) and $1.6\times 10^{-8}$ 
($6.2\times 10^{-10}$), respectively. Comparing with the purely leptonic 
decays of $B_{s,d} \to l^+ l^-$, we have shown that the  branching ratios 
of $B_{s,d} \to l^+ l^- \gamma$ have the same order 
of magnitude for the $\mu$ channel but that of $B_{s,d} \to e^+ e^-
\gamma$ 
are much larger.
We conclude that some of the radiative leptonic decays of
$B_{s,d} \to \l^+ \l^- \gamma$ could be measured in the $B$ factories
as well as LHC-B experiments, where approximately,
$6 \times 10^{11}(2 \times 10^{11})~B_d(B_s)$ mesons are expected to be
produced per year.\\


\vspace{1cm}

\noindent
{\bf Acknowledgments}

This work was supported in part by the National Science Council of the
Republic of China under contract number NSC-89-2112-M-007-013 and
NSC-89-2112-M-006-026 .

\newpage

\begin{figcap}

\item
The values of the form factors $F_{TA}$ (solid curve) and $F_{TV}$ (dashed
curve) as functions of the momentum transfer $p^2$
for $B_s\to \gamma$.
\item
The differential decay
branching ratio $dB(B_{s}\to \mu^+ \mu^- \gamma)/dx_{\gamma}$ as a
function of
$x_{\gamma}=2E_{\gamma}/M_{B_{s}}$ with(solid curve) and without (dashed
curve) long distance.
\item
The differential decay
branching ratio  $dB(B_{s}\to \tau^+ \tau^- \gamma)/dx_{\gamma}$ as a
function of
$x_{\gamma}=2E_{\gamma}/M_{B_{s}}$ with (solid curve) and without (dashed
curve) long distance.

\end{figcap}

\newpage
\begin{figure}[h]
\includegraphics{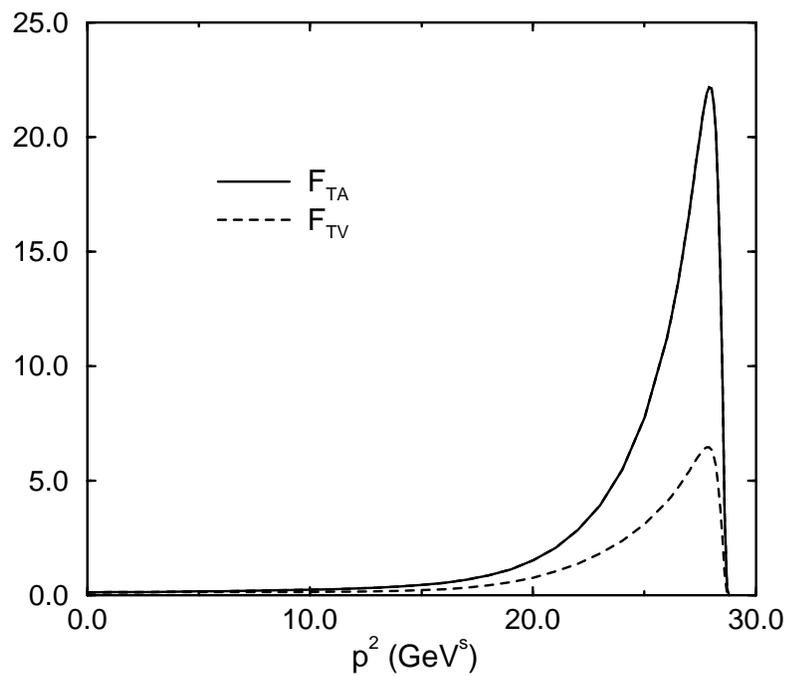}
\vskip 13cm
\caption{
The values of the form factors $F_{TA}$ (solid curve) and $F_{TV}$ (dashed
curve) as functions of the momentum transfer $p^2$
for $B_s\to \gamma$.
}
\end{figure}

\newpage
\begin{figure}[h]
\includegraphics{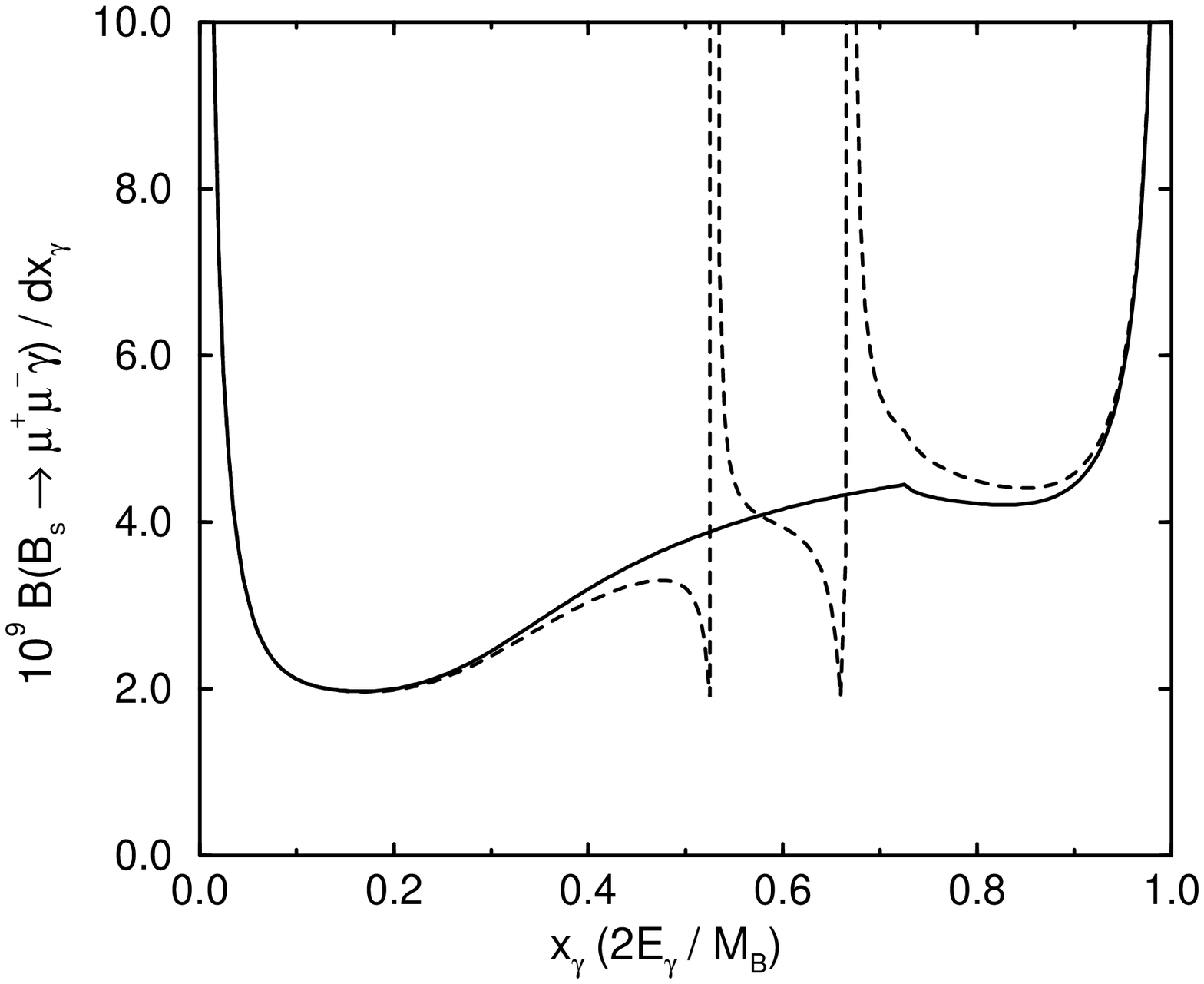}
\vskip 13cm
\caption{The differential decay
branching ratio $dB(B_{s}\to \mu^+ \mu^- \gamma)/dx_{\gamma}$ as a
function of
$x_{\gamma}=2E_{\gamma}/M_{B_{s}}$ with (solid curve) and without (dashed
curve) long distance.}
\end{figure}

\newpage
\begin{figure}[h]
\includegraphics{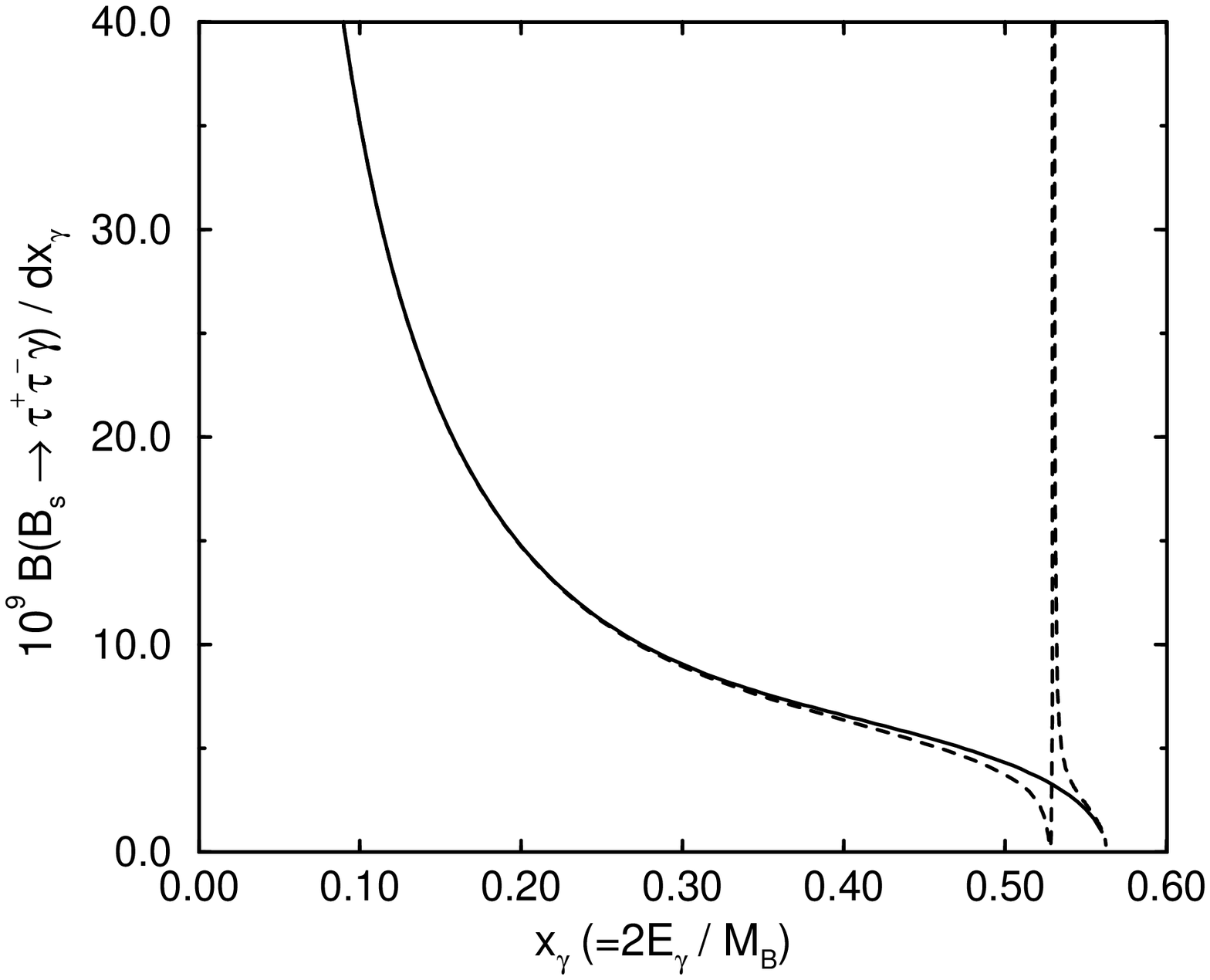}
\vskip 13cm
\caption{The differential decay
branching ratio $dB(B_{s}\to \tau^+ \tau^- \gamma)/dx_{\gamma}$ as a
function of
$x_{\gamma}=2E_{\gamma}/M_{B_{s}}$ with (solid curve) and without (dashed
curve) long distance.}
\end{figure}

\end{document}